\documentstyle[epsfig]{mn}

\title[Dwarf spheroidal star formation histories]{Non-parametric star formation histories
for 5 dwarf spheroidal galaxies of the local group}
\author[X. Hernandez, G. Gilmore and D. Valls-Gabaud]
{X. Hernandez$^{1,2}$, Gerard Gilmore$^1$ and
David Valls-Gabaud$^{3,1}$ \\
$^1$ Institute of Astronomy, Cambridge University, Madingley Road, Cambridge CB3 0HA \\
$^2$ Instituto de Astronom\'\i a, Universidad Nacional Aut\'onoma de M\'exico, A.P. 70-264, 
04510 M\'exico, D.F. \\
$^3$ UMR CNRS 7550, Observatoire de Strasbourg, 11 Rue de l'Universit\'e, 67000 Strasbourg, France. \\
}

\date{\today}

\begin{document}
\maketitle

\begin{abstract}

We use recent HST colour-magnitude diagrams of the resolved stellar populations of a 
sample of local dSph galaxies (Carina, LeoI, LeoII, Ursa Minor and Draco) to infer the 
star formation histories of these systems, $SFR(t)$. Applying a new variational calculus maximum 
likelihood method which includes a full Bayesian analysis and allows a non-parametric
estimate of the function one is solving for, we infer the star formation histories of the systems studied. 
This method has the advantage of yielding an objective answer, as one need not assume {\it a priori}
the form of the function one is trying to recover. The results are checked independently using Saha's 
$W$ statistic. The total luminosities of the systems are used
to normalize the results into physical units and derive SN type II rates. We derive the luminosity weighted mean star 
formation history of this sample of galaxies.

\end{abstract}


\section{Introduction}

The local dwarf spheroidal galaxies form a sample of small galaxies which, due to their 
relatively nearby locations and close association with the Milky Way could in principle furnish 
crucial observational and theoretical information on a range of astrophysical phenomena.
As these systems are eventually disrupted and incorporated into the Milky Way they illustrate 
locally one of the mechanisms thought to be responsible for the build up of large galaxies. 
Thus comparing their stars and those now found in our galaxy we can obtain a first estimate of 
the relevance of late mergers (Unavane et al. 1996). Dynamical studies of their stars have yielded 
valuable constraints on the nature and structure of dark matter halos at the smallest scales 
(Lin \& Faber 1983, Gerhard \& Spergel 1992). Their orbits probe the galactic halo in a range of 
distances not sampled by any other objects and can thus be used to study the outer galactic halo. 
Additionally, their small sizes make them in principle the simplest galactic systems, where key 
processes such as star formation and gas flows can be studied under relatively well defined 
conditions. 

However, this situation is complicated by the fact that we not only lack a theoretical
understanding of these systems, but also an observational record of their evolution; only a present
day snapshot of their physical parameters is available, as is the case with most galactic systems. 
Whilst high redshift observations have recently opened up new areas of research as they begin to 
yield an statistical description of the evolution of bright galaxies, such an approach is likely to 
remain out of reach for these small systems for some time. Fortunately, their neighboring locations
allow the study of their individual stars, which offers the possibility of directly probing their 
evolutionary histories by inferring star formation rates as a function of time, $SFR(t)'s$.

The recent availability of detailed colour-magnitude diagrams for several nearby systems has
prompted the development and allowed the application of careful statistical methods aimed at
reconstructing the star formation histories of these objects (e.g. Chiosi et al. (1989), 
Aparicio et al. (1990) and Mould et al. (1997) using Magellanic
and local clusters, and Mighell \& Butcher (1992), Smecker-Hane et al. (1994), Tolstoy \& Saha 
(1996), 
Aparicio \& Gallart (1995) and Mighell (1997) using local dSph's). Although much has been learnt 
of the complex $SFR(t)'s$ of these systems, existing studies have lacked two major ingredients: 
a homogeneous set of observations including several of the dSph galaxies does not exist, and 
different data sets are generally analyzed using different techniques. These two points make 
comparisons between the derived $SFR(t)'s$ at best risky. A further difficulty lies in the fact 
that the available rigorous statistical studies approach the problem parametrically, which is 
something one should try to avoid when the actual structure of the function one is trying to 
recover can be crucial, as is the case when the underlying physics is unknown. An example of this 
last point is the case of the Carina dwarf. Hurley-Keller et al. (1998) solve for the best fitting 
three discrete bursts solution to the $SFR(t)$ and conclude that star formation has proceeded 
spasmodically, whilst Mighell (1997) uses a non parametric star count approach, albeit not a 
fully consistent statistical method, to obtain a more gradual solution for Carina's $SFR(t)$.
	
In this paper we have attempted to improve on the determination of the star
formation histories of local dSph systems by addressing the two points mentioned above. 
We use recent HST observations of the resolved populations
of a sample of dSph galaxies (Carina, LeoI, LeoII, Ursa Minor and Draco) uniformly taken 
and reduced, to recover the $SFR(t)$ of each, applying a new non-parametric maximum 
likelihood method. This allows meaningful comparisons to be made, as any systematics, at any
level, will affect all our galaxies equally.

The outline of our paper is as follows: in section 2 we discuss the observations, in section
 3 we include a brief outline of our method, which was introduced in our paper I (Hernandez 
et al. 1999).
The results are presented in section 4, and in section 5 we summarize our results.

\section{The observations}

The main requirements of our observations were that they should comprise a homogeneous 
sample of local dSph galaxies, mostly in terms of the data reduction. Only such an internally
consistent data set allows robust comparisons between different galaxies, once uniform
data reduction and analysis methods are adopted. We extracted available archive HST data for
the Carina, LeoI, LeoII, Ursa Minor and Draco galaxies, and used standard data reduction 
methods
and standard HST calibration numbers throughout the sample (e.g. Elson et al. 1996).

The currently available data cover only small (and variable) sections of the total extent of 
these systems. This fact clearly limits the inferences which can be drawn to the small
observed fractions, the star formation histories of these regions might not by representative
of the average for a whole galaxy. While this limitation introduces an extra uncertainty
to our results, it highlights the interesting possibility of studying spatial variations in the 
evolutionary histories of dSph galaxies, if comprehensive HST studies where undertaken. 
In the above sense, our results for the different galaxies refer in the strictest sense only
to the fractions covered by the observed fields.

As we did not require the HR diagrams to extend much fainter than the oldest turnoff points
($M_V \approx 24 - 25$), or to be complete into the faintest limits (the faintest stars 
were in fact
excluded from the analysis) the data reduction was straightforward. Our resulting CMDs do not
show any systematic difference from comparable published ones for the galaxies we study. The
technical details of the images used appear in the appendix.

\section{The method}

In this section we give a summary description of our HR diagram inversion method, which was
described extensively in our paper I. In contrast with other statistical methods, we do not
need to construct synthetic colour magnitude diagrams (CMD) for each of the possible 
star formation histories being considered. Rather we use a direct approach which solves 
for the best 
$SFR(t)$ compatible with the stellar evolutionary models assumed and the observations used. 

The evolutionary model consists of an isochrone library,
and an IMF. Our results are largely insensitive to the details of the latter, for which 
we use:

\begin{equation}
\rho(m) \propto \left\{
\begin{array}{rl}
     m^{-1.3} &  0.08M_{\odot} <m\le 0.5 M_{\odot} \\[1.0 ex] 
     m^{-2.2} &  0.5M_{\odot} <m\le 1.0 M_{\odot} \\[1.0 ex]
     m^{-2.7} &  1.0M_{\odot} < m
\end{array}\right.
\end{equation} 
The above fit was derived by Kroupa et al. (1993) for a large sample towards both galactic 
poles and
all the solar neighborhood.

As the weak metallicity dispersions measured in the galaxies we are studying (around 0.3
 dex) are comparable to the errors in the metallicity determinations themselves, we have 
not attempted to introduce
any enrichment histories for any of our galaxies. In fact, as small internal metallicity 
spreads present in these galaxies would introduce only small differential age offsets in our
inferred $SFR(t)$ (see Table I), we shall in all cases use single metallicity isochrone 
sets. Once a metallicity has been selected, we use the latest Padova isochrones (Fagotto et 
al. 1994, Girardi et al. 1996) together with a detailed constant phase interpolation scheme 
using only stars at constant evolutionary phase, to construct an isochrone library having a 
chosen temporal resolution. 

In this case we implement the method with a resolution of 0.15 Gyr, sufficient for our 
present problem. It is one of the advantages of the method that this resolution can be 
increased arbitrarily (up to the stellar model resolution) with computation times scaling 
only linearly with it. 

Our only other inputs are the positions of, say $n$ observed stars in the HR diagram, each 
having a colour and luminosity, $c_{i}$ and $l_{i}$. Starting from a full likelihood model, 
we first construct the probability that the $n$ observed stars resulted from a certain 
$SFR(t)$. This will be given by:

\begin{equation}
{\cal L}= \prod_{i=1}^{n} \left( 
\int_{t_0} ^{t_1} SFR(t) G_{i}(t) dt \right),
\end{equation}

where
$$
G_{i}(t)= {\rho(l_i;t) \over{\sqrt{2 \pi} \sigma(l_i)}} 
exp\left(-\left[C(l_i;t)-c_{i}\right]^2 \over {2 \sigma^2(l_i)} \right)
$$

In the above expression $\rho(l_i;t)$ is the density of points along the isochrone
of age $t$, around the luminosity of star $i$, and  is determined by the assumed IMF 
together with the duration of the differential phase around the luminosity of star $i$. 
$t_0$ and $t_1$ are a maximum and a minimum time needed to be
considered, for example 0 and 15 Gyr. $\sigma(l_i)$ is the amplitude of the 
observational errors in the colour of the stars, which are a function of the luminosity
of the stars. This function is supplied by the particular observational sample one is
analyzing. Finally, $C(l_i;t)$ is the colour the observed star would actually have if it had
formed at time $t$.  We shall refer to $G_{i}(t)$ as the
likelihood matrix, since each element represents the probability that a given star, $i$,
was actually formed at time $t$. Since the colour of a star having a given luminosity and
age can sometimes be multi-valued function, in practice we check along a given isochrone, to
find all possible masses a given observed star might have as a function of time, and add all
contributions (mostly 1, sometimes 2 and occasionally 3) in the same $G_{i}(t)$.
In this construction we are only considering observational
errors in the colour, and not in the luminosity of the stars. The generalization
to a two dimensional error ellipsoid is trivial, however the observational errors in 
colour dominate 
the problem to the extent of making this refinement unnecessary. Although the amplitude of 
luminosity errors is only a factor of $\approx 2$ smaller than colour errors, as can be 
inferred from the fact that CMD diagrams typically display a range of luminosities 5 times 
larger than in colour, in discriminating between isochrones, errors in colour are 
$\approx 10$ times as important as errors in luminosity.
The absence of a colour dependence from $\rho(l_i;t)$ is a direct consequence of having
neglected errors in the luminosity of the stars. A star of a given observed luminosity and
assumed age will thus have a colour determined by the isochrones used. 

Equation (2) is essentially the extension from the case of a discrete $SFR(t)$ used 
by Tolstoy \& Saha (1996), to the case of a continuous function (continuous in time, 
but obviously discrete with respect to the stars) in the construction of the likelihood. 
The challenge now is to find the optimum $SFR(t)$ without evaluating equation (2) i.e.
without introducing a fixed set of test $SFR(t)$ cases from which one is selected. 
 
The condition that ${\cal L}(SFR)$ has an extremal can be written as
$$
\delta {\cal L}(SFR)=0,
$$
and a variational calculus treatment of the problem applied. 
Firstly, we develop the product over $i$ using the chain
rule for the variational derivative, and divide the resulting sum by ${\cal L}$ to obtain:

\begin{equation}
\sum_{i=1}^{n} \left(
{\delta \int_{t_0} ^{t_1} SFR(t) G_{i}(t) dt} \over {\int_{t_0}^{t_1} SFR(t) G_{i}(t) dt}
\right) =0
\end{equation}

Introducing the new variable $Y(t)$ defined as:

$$
Y(t)=\int{ \sqrt {SFR(t)} dt} \Longrightarrow  SFR(t)=\left( {dY(t)
\over dt} \right)^2
$$

and introducing the above expression into
equation~(3) we can develop the Euler equation to yield, 

\begin{equation}
{d^2 Y(t)\over dt^2}\sum_{i=1}^{n} \left( G_{i}(t) \over I(i)\right)
=-{dY(t)\over dt}\sum_{i=1}^{n} \left( dG_{i}/dt \over I(i)\right)
\end{equation}

where 
$$
I(i)=\int_{t_0}^{t_1} SFR(t) G_{i}(t) dt
$$

This in effect has transformed the problem from one of searching for a function
which maximizes a product of integrals (equation 2) to one of solving
an integro-differential equation (equation 4).  We solve this equation iteratively,
with the boundary condition SFR(15)=0. Details of the numerical procedure required to 
ensure convergence to the maximum likelihood SFR(t) can be found in our paper I, where the
method is tested extensively using synthetic HR diagrams.
The main advantages of our method over other maximum likelihood schemes are the totally
non parametric approach the variational calculus treatment allows, and the efficient computational
procedure, where no time consuming repeated comparisons between synthetic and observational 
CMD are necessary, as the optimal $SFR(t)$ is solved for directly.

The lower main sequence region of the CMD diagram is totally degenerate with age, and 
contains the lower brightness stars, where the errors are larger. We have seen from using
synthetic HR diagrams that excluding this region produces a faster and more accurate convergence of
the method, and have in analyzing real galaxies excluded stars of magnitudes fainter than 
$M_V \approx +5$. This
last cut together with the fact that our isochrones only extend out to the tip of the red giant
branch (to go further would necessitate combining results from different physical models, which we
preferred not to do) leaves us with a mass range which actually varies as a function of time. 
To include also the fraction of the $SFR(t)$ outside this region, we apply a minor correction
factor to the result of equation (4), which accounts for the fraction of mass outside the
sampled range, as a function of time, as given by the IMF used.

\begin{table}
 \caption{Age offset in Gyr between simulated and recovered populations
 as a function of input age (in Gyr) and metallicity (in solar units), for a $\pm$ 0.2 dex metallicity mismatch}
 \label{symbols}
 \begin{tabular}{|l|c|c|c|c|c|}
	\hline
  $[Fe/H]$  & -1.4 & -1.6 & -1.8 & -2.0 & -2.2 \\ \hline

  2 Gyr & +0.6 & +0.4 & +0.3 & +0.4 &      \\ 
    &      & -0.3 & -0.1 & -0.2 & -0.2 \\ \hline
  
  4 Gyr & +0.8 & +0.5 & +0.4 & +0.6 &      \\ 
    &      & -0.6 & -0.4 & -0.2 & -0.4 \\ \hline

  6 Gyr & +1.6 & +0.5 & +0.7 & +0.9 &      \\ 
    &      & -0.9 & -0.2 & -0.5& -0.6  \\ \hline

  8 Gyr & +1.8 & +0.4 & +0.5 & +1.4 &      \\ 
    &      & -1.3 & -0.8 & -0.6 & -0.9 \\ \hline

 10 Gyr & +2.5 & +0.5 & +0.7 & +1.3 &      \\ 
    &      & -2.0 & -0.5 & -0.7 & -1.5 \\ \hline

\end{tabular} 

\end{table}

Before presenting the star formation histories which result from applying our method to the colour-magnitude
diagrams of the galaxies sampled, we include a summary of the systematic errors associated 
with the theoretical inputs of the method, which are more extensively discussed in our paper I.

The IMF convolved with the duration of the differential evolutionary phase enters the calculation 
of the likelihood matrix in determining the density of points around the luminosity of each 
observed star, for each of the isochrones considered. As the main sequence and the giant branch
regions of the CMD are degenerate with age (for a single metallicity population), it is the region 
containing the turn-off points of the sampled population that drives the solution of the problem.
As a consequence of the above, the details of the IMF used are largely unimportant, it is basically
the main sequence lifetime of a star that the solution is sensitive to. This was shown explicitly 
in our paper I through the use of synthetic HR diagrams, where the IMF affected only the amplitude
of the recovered $SFR(t)'s$, which were normalized through the total number of stars in the HR 
diagrams. As in this case we are normalizing the inferred $SFR(t)'s$ through the total luminosities
of the galaxies being studied, changing the IMF within any reasonable limits leaves the results 
unaffected. The effect of any blended binaries is equally unimportant, as the broadening of the 
main sequence occurs in the degenerate region and is in any case much smaller than the broadening 
produced by the observational colour spread. 

The resolution of the method varies as a function of time and of the observational errors present in
the CMD being analyzed, in ways that were studied in our paper I. The observational errors tend to 
smear the time structure in the $SFR(t)$ always towards older ages, i.e. a burst of age $t$ will be
recovered as an episode of duration $\Delta t$ ending at time $t$. This  $\Delta t$ varies with 
age, becoming significant $(>1 Gyr)$ only for populations older than around 10 Gyr, for the level 
of observational errors present in our HST colour-magnitude diagrams. Younger populations are less 
affected, and the formal resolution at which the method was implemented of 0.15 Gyr is 
representative of our results for ages younger than 6 Gyr.

As explained in the previous section, our isochrones end at the tip of the red giant branch, 
which means that more advanced evolutionary phases can not be incorporated into the analysis. 
Fortunately, these later phases occupy regions of the CMD diagram distinct from those containing 
the phases we account for. Therefore, we can simply remove any red clump and horizontal branch 
stars from the analysis, leaving the structure of the studied regions unaffected. These later 
phases form only a minority component, containing little extra information, and do not affect our 
inferences. In the same way our results are not affected by the presence of some contaminating 
field stars. Provided they do not fall on the region of the CMD containing
the turn off points of the underlying stars, they are simply removed from the analysis. 

We are however sensitive to the assumed metallicity of the stellar populations being treated. In our
paper I we presented a few examples of how the method reacts when inverting a synthetic HR diagram 
produced with isochrones different to the ones used in the inference procedure.  If we construct 
the likelihood matrix using isochrones which are very different (about 1 dex) from the ones used 
to produce the HR diagram, the iterative method used in solving equation (4) becomes unstable and 
tends to divergent solutions. This property can be used to deduce large incompatibilities between 
the stars and the template isochrones against which they are being compared. Small metallicity 
offsets are much harder to detect, and produce distorted results. The degree of distortion varies 
with age, shape of the overall $SFR(t)$ and the observational errors present in highly non-linear 
fashion. 

To give some indication of these distortions we present Table I. We produced a
synthetic HR diagram from a single Gaussian burst input $SFR(t)$ having a duration of 1 Gyr, of a 
given metallicity, and applied the method using a metallicity 0.2 dex lower than the one used to
generate the stars. This was repeated for a range of metallicities and ages, for both positive 
and negative metallicity mismatches. The top row of Table 1 shows the input metallicity (in dex), 
the first column shows the input burst age, and the other entries show the age offset between the 
recovered and input $SFR(t)$, all ages are given in Gyr. A ``+'' sign denotes the inferred 
population was older than the input one, where a lower metallicity was used in the inference 
procedure. Similarly, a ``-'' indicates that the inferred population was younger than the input 
one, where a higher metallicity was used to invert the simulated CMD.

The metallicities shown 
cover the range present in the galaxies being studied. Populations much younger than 2 Gyr are 
not present in our galaxies, and those older than 10 Gyr are distorted by the observational errors 
to the point of eliminating much of the time resolution of the inversion in this region. This 
table can be used to estimate the effects of changing the assumed metallicities, or of introducing 
temporal gradients, within the small observationally restricted range.
 The well known age-metallicity
degeneracy is apparent. This might affect our results even if the mean metallicity is well known, as
temporal variations in the metallicities (which must exist at some level) are not considered 
by the method. Using an independent test on our results, we find these effects to be minor in most
of the cases we study, as observational measurements of the metallicity of these systems suggest.

\begin{figure*}
\epsfig{file=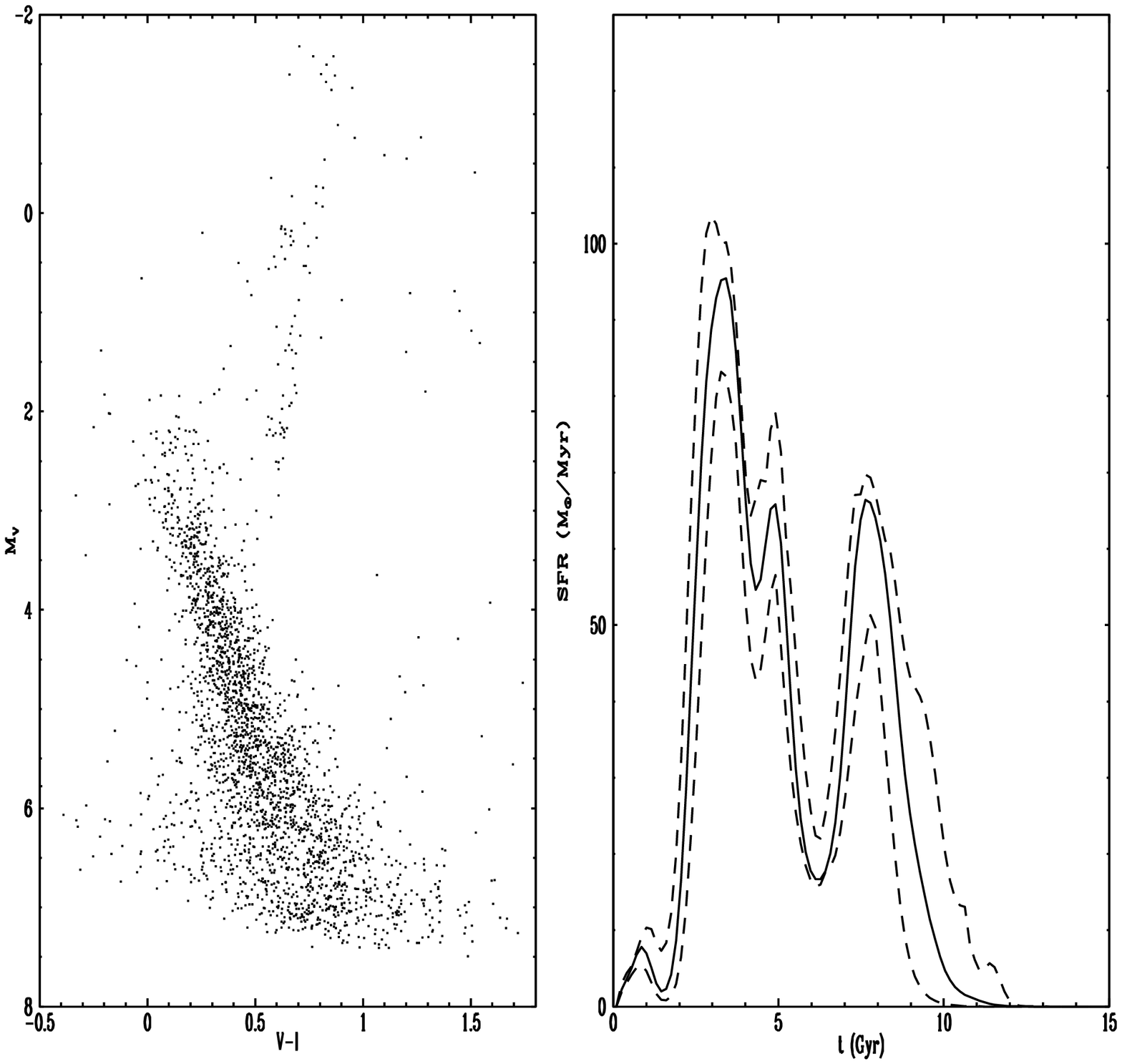,angle=0,width=18.1cm,height=9.0cm}
\@ \textbf{Figure 1.}\hspace{5pt}{
\begin{flushleft}\textbf{Left:} Observational HR diagram for Carina.
\textbf{Right:} Inferred $SFR(t)$ for the central values of the observational parameters, solid 
line. The dashed curves represent the error envelope as defined by the quoted uncertainties in 
the foreground extinction and distance modulus. 
\end{flushleft}}
\end{figure*}

\subsection{Testing the results}

Once the IMF, metallicity and observational parameters are assumed for a given galaxy, the positions
of the observed stars in the CMD are used to construct the likelihood matrix $G_{i}(t)$,
which is the only input given to the inversion method. In our Paper I we tested this method using 
synthetic CMDs produced from known $SFR(t)'s$, with which we could assess the accuracy of the result
of the inversion procedure. In working with real data, we require the introduction of an independent
method of comparing our final result to the starting CMD, in order to check that the answer our
inversion procedure gives is a good answer. From our paper I we know that when the stars being used
in the inversion procedure were indeed produced from the isochrones and metallicity used to 
construct the likelihood matrix, the inversion method gives accurate results. The introduction of 
an independent comparison between our answer and the data is hence a way of checking the accuracy 
of the input physics used in the inversion procedure, i.e. the IMF, metallicity and observational 
parameters.

The most common procedure of comparing a certain $SFR(t)$ with an observed CMD is to use the 
$SFR(t)$ to generate a synthetic CMD, and compare this to the observations using a statistical test
to determine the degree of similarity between the two. For example, Aparicio et al. (1997) 
manage to recover simultaneously the distance, enrichment history and $SFR(t)$ of the local dwarf 
LGS 3 by constructing synthetic CMD's from a $SFR(t)$ taken parametrically as a series of 
contiguous bursts, and finding the amplitudes of each burst that give the maximum statistical 
similarity with the data, in terms of a counts in cells maximum likelihood. In that case, they 
solve for the amplitudes of the bursts that give a total synthetic CMD most closely resembling 
the observational data set. Their synthetic CMD is a linear sum of the partial CMD's produced from 
a single realization for each burst. This has the advantage of allowing a large parameter space to 
be considered, as the synthetic CMD's are constructed trivially from a fixed single statistical 
realization of each burst. The disadvantage however is that one is not comparing the $SFR(t)$ with 
the data, but rather a particular realization of the $SFR(t)$ with the data. The distinction
becomes arbitrary when large numbers of stars are found in all regions of the CMD, which is 
generally not the case. Following a Bayesian approach, we prefer to adopt the $W$ statistic 
presented by Saha (1998), essentially

$$
W=\prod_{i=1}^{B} {{(m_{i}+s_{i})!} \over {m_{i}!s_{i}!}}
$$

where B is the number of cells into which the CMD is split, and $m_{i}$ and $s_{i}$ are the 
numbers of points two distributions being compared have in each cell. This asks for the probability
that two distinct data sets are random realizations of the same underling distribution. In 
implementing this test we first produce a large number ($\sim 500$) of random realizations of our 
best answer $SFR(t)$, and compute the $W$ statistic between pairs in this sample of CMD's. This 
gives a distribution which is used to determine a range of values of $W$ which are expected to 
arise in random realizations of the $SFR(t)$ being tested. Next the $W$ statistic is computed 
between the observed data set, and a new large number of random realizations of $SFR(t)$, this 
gives a new distribution of $W$ which can be objectively compared to the one arising from the 
model-model comparison to assess whether both data and modeled CMD's are compatible with a unique 
underling distribution. Both distributions of $W$ were characterized in terms of a mean value 
and a $1 \sigma$ amplitude. This final check of our answer is in fact the slowest part of the 
procedure, but necessary to obtain an independent check on the answer of our inversion method. 
In other terms, we are checking that our best inferred maximum-likelihood solution is also a 
good fit. The value of B used was $\sim$ 6400.

\section{The galaxies}

\subsection{Carina}

The first galaxy we study is the Carina dwarf, one of the first dSph's to be observed in terms
of resolved stellar population, and the one for which the most studies inferring the $SFR(t)$ 
from the CMD  have been published. This gives us the opportunity of comparing our results with
previous studies. Our CMD diagram initially contained 2550 stars. After removing contamination, stars
beyond the RGB and the lower degenerate region, we are left with 980 stars. The full observational
CMD is shown in the left panel of Figure (1). 

\begin{figure*}
\epsfig{file=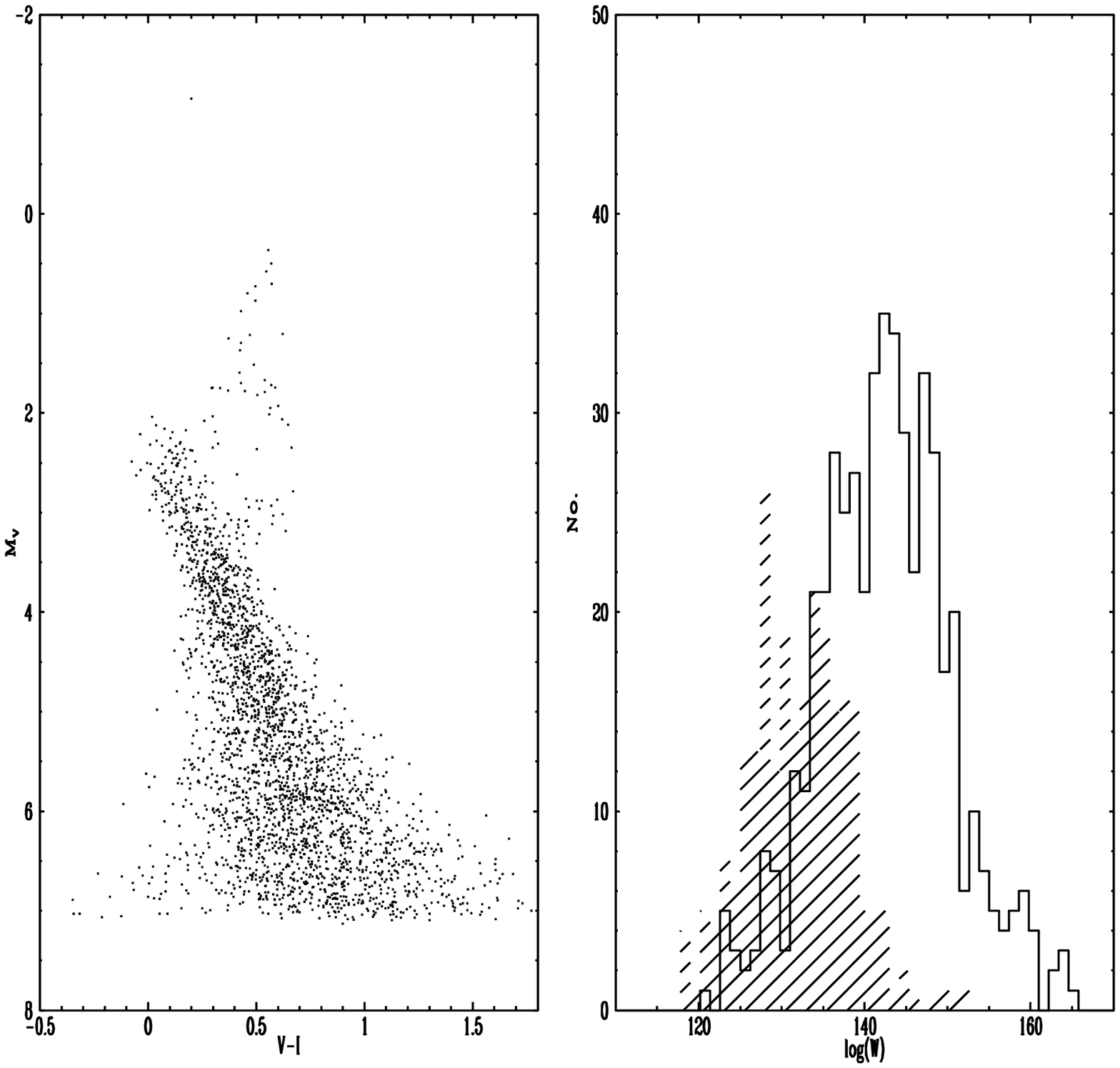,angle=0,width=18.1cm,height=9.0cm}
\@ \textbf{Figure 2.}\hspace{5pt}{
\begin{flushleft}\textbf{Left:} One synthetic HR diagram for Carina produced using the central 
inferred $SFR(t)$ of Figure (1).
\textbf{Right:} Distribution of values of W arising from 500 model-model comparisons, unshaded
histogram. This establishes the range of values of W which result from random realizations of
the inferred $SFR(t)$. Also shown is the distribution of values of W arising from 250 data-model
comparisons, showing the hypothesis that the data result from the same underlying function as the
models to be acceptable at more than 1 sigma level. 
\end{flushleft}}
\end{figure*} 

Using the numbers published in the recent review by Mateo (1998) we took as the central values
of our observational parameters for Carina $[Fe/H]=-2.0 \pm 0.2$, $M_{V}=-9.3$, 
$E(B-V)=0.04\pm0.02$ and $(m-M)_{0}=20.03 \pm0.09$ (Smecker-Hane et al. 1994, Mighel 1997, 
Hurley-Keller et al. 1998 and Mateo et al. 1998). The metallicity fixes the isochrones and the 
integrated magnitude the normalization, we use the distance modulus and the reddening correction 
to fix the observations in the CMD. At this point we apply our method to invert the observational 
CMD and recover the underlying $SFR(t)$, this is shown in the right panel of Figure (1) by the 
solid curve. The dotted curves represent the upper and lower envelopes to a series of alternative 
reconstructions of the $SFR(t)$ produced by changing the assumed values of the distance modulus 
and the reddening corrections, within their respective error ranges. The internal metallicity 
spread present in this galaxy is very low, as can be seen from the narrow RGB, and is quoted as 
$<0.1$ dex in the review by Mateo (1998). In this way, our error margins for this galaxy due to 
metallicity uncertainties are not much larger than what is shown by the dotted lines, see Table(1). 

In Figure (2) we illustrate the procedure of checking the inferred $SFR(t)$, the left panel gives
one random realization of the central inferred $SFR(t)$, which is seen to resemble the data
for Carina rather well. The right panel of Figure (2) shows the implementation of the W test.
The solid histogram gives the distribution of values of W which arise from 500 model-model
comparisons, and gives the variability arising from the different random realizations of the 
central $SFR(t)$, for the number of stars present in our observations. The dashed histogram
gives the distribution of values of W which result from 250 data-model comparisons. Only
$<32$ \% of the random realizations of our central $SFR(t)$ would give distributions of W
(when compared against all other realizations) having a mean value further removed from that of
the 500 model-model distribution than the data. In this sense, we can accept the hypothesis that
both the data and our 500 random realizations of the central $SFR(t)$ for Carina come from
the same underlying generating function at a 1 $\sigma$ level. For the remaining galaxies we 
shall give only the results of the W test in terms of the mean and 1$\sigma$ amplitude of the 
model-model and the data-model distributions.

Our result shows an interesting $SFR(t)$ for this galaxy, very little star formation at early times,
until around 10 Gyr ago, when over a period of 3 Gyr an intermediate population was formed.
The $SFR(t)$ then decreased markedly, before entering a more recent and extended period of star
formation which ended 2 Gyr ago. The very low levels found throughout
for the star formation rate of $50-100 M_{\odot}/Myr$ are representative of what we find in all 
our galaxies, and should provide clues as to the physical processes driving the star formation 
activity in these systems.

\begin{figure*}
\epsfig{file=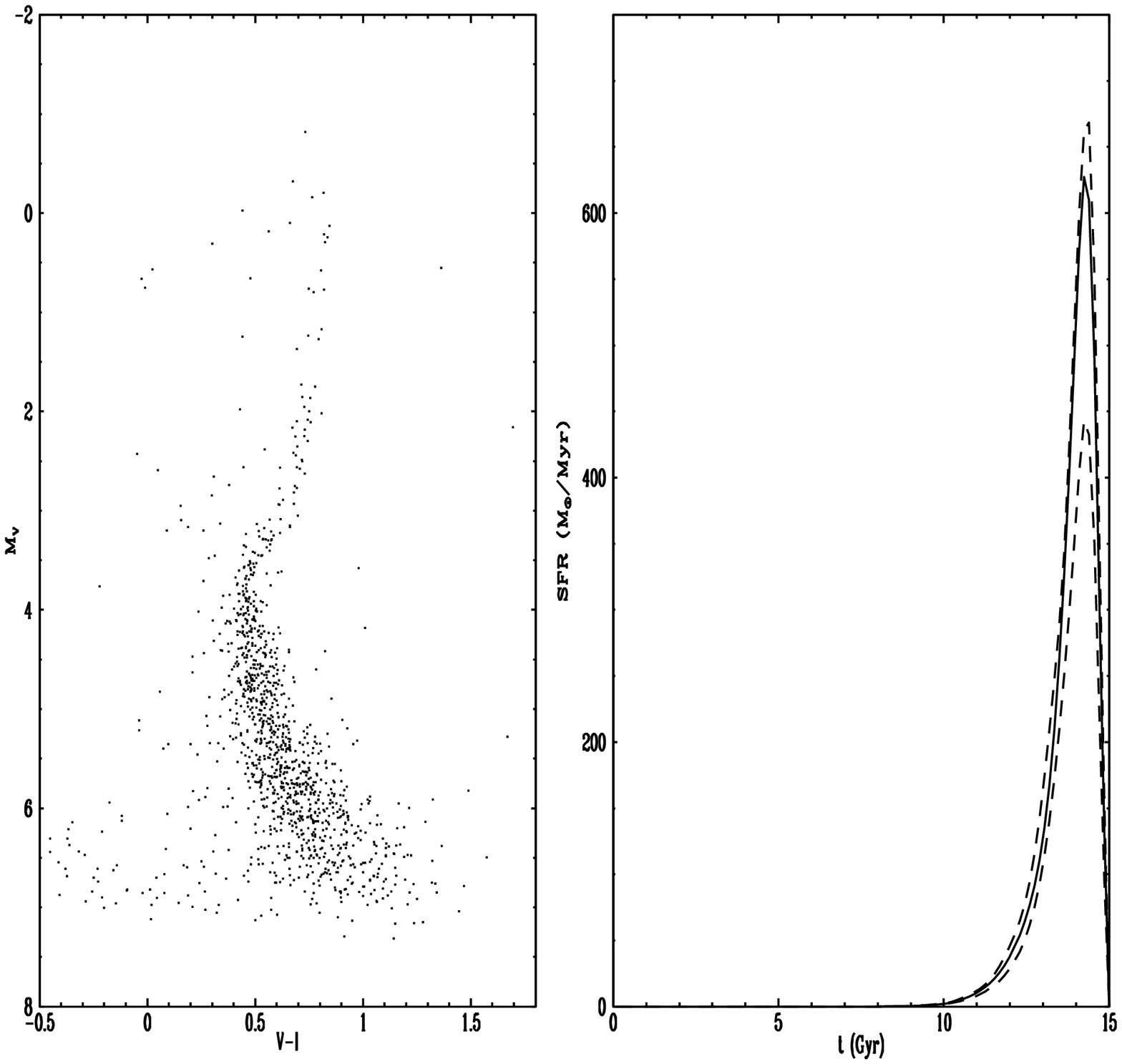,angle=0,width=18.1cm,height=9.0cm}
\@ \textbf{Figure 3.}\hspace{5pt}{
\begin{flushleft}\textbf{Left:}  Observational HR diagram for Ursa Minor. 
\textbf{Right:}  Inferred $SFR(t)$ for the central values of the observational parameters, 
solid line. The dashed curves represent the error envelope as defined by the quoted 
uncertainties in the foreground extinction and distance modulus. 
\end{flushleft}}
\end{figure*}

The existence of some RR Lyr stars in this galaxy (e.g. Saha et al. 1986, Mateo et al. 1995 or
Kuhn et al. 1996)
signal the presence of a very old population at some level, although recent estimates of the 
$SFR(t)$ in Carina coincide in that the amplitude of this old component is minor
(Hurley-Keller \& Mateo 1998), and appears to have blended completely into the MS of our CMD, 
or to be found preferentially in a region of the galaxy not sampled by our observations. One 
of the most general claims about the $SFR(t)$ of Carina has been the extreme ``bursting'' 
character of it (e.g. Smecker-Hane 1994, Hurley-Keller \& Mateo 1998) we note however, that all 
these studies have assumed {\it a priori} an extremely discrete form for the $SFR(t)$ of this 
galaxy, and then solved for the best such function. In contrast, Mighell (1997) using a 
non-parametric approach, finds a much more continuous solution, basically consistent with what 
we obtain. The resolution of our method in the region of ages $2-8 Gyr$ is sufficient to exclude 
the possibility of any total cessation of the star forming activity in this region lasting more 
than $0.5 Gyr$, as was shown in our Paper I, where synthetic HR diagrams produced from bursting 
$SFR(t)'s$ were correctly inverted. We conclude that although the star formation history of this 
galaxy is clearly bimodal, it is not a series of discrete bursts lasting $\le 1 Gyr$. However,
we can not exclude the possibility that sampling a much larger fraction of this galaxy could
yield somewhat different results. Analyzing local variations of the $SFR(t)$ within these galaxies 
is an interesting project which will be treated in other papers, using data covering greater 
portions of the galaxies. As with all our galaxies, given the small number of stars available,
the resolution of our solution is limited. Episodes of very short duration or low level, 
resulting in very few stars will be totally missed. We are recovering the $SFR(t)$ responsible
for producing the greater fraction of the observed stars.

\subsection{Ursa Minor}

The case of Ursa Minor seems to be the simplest of the ones we study, and is actually the only 
one of our galaxies which agrees with the once common expectation of dSph systems being simply old 
and metal poor. Our observational CMD is shown in the left panel of Figure (3), and is made up of 
1232 stars. After removing those stars incompatible with the phases included in our isochrones, 
together with the fraction beneath $M_{V}=6$ we are left with 334 stars, which we used in the 
inversion procedure.

Using the values given in Mateo's (1998) review, we take $[Fe/H]=-2.2 \pm0.1, M_{V}=-8.9, 
E(B-V)=0.03\pm0.02$ and $(m-M)_{0}=19.11 \pm0.1$ (Nemec et al. 1988 and Olszewski \& 
Aaronson 1985) for this galaxy. The reported internal metallicity
dispersion in this galaxy is also very low, at $<0.2$ dex, which introduces little uncertainty in 
our results. Applying our method using the central values of the observational parameters we obtain
the solid line shown in the right panel of Figure (3). Again, the dotted curves represent an 
envelope to a large number of reconstructions obtained by changing the observational parameters 
within their error ranges. Of this galaxy we can say that most of its stars are older than 12 Gyr. 
Given the observational errors present and the large age of the population of this galaxy, we can 
not draw any inferences on the time structure of the $SFR(t)$, as this is totally lost in the 
noise. The duration of the star forming episodes can only be concluded to have been $\le 3 Gyr$. 
Normalizing through the total luminosity of this galaxy we obtain rates of $>400 M_{\odot}/Myr$. 
The result of applying the $W$ test to this galaxy gives $47 \pm 5$ and $44 \pm 4$ for the 
model-model and model-data sets, respectively, showing our answer to be compatible with the data 
at better than a 1 $\sigma$ level. Olszewski \& Aaronson (1985) used a ground based CMD and a 
simple isochrone fitting procedure to conclude the population of this galaxy is uniformly old, 
with the possibility of a 2Gyr spread.

\subsection{LeoI}
\begin{figure*}
\epsfig{file=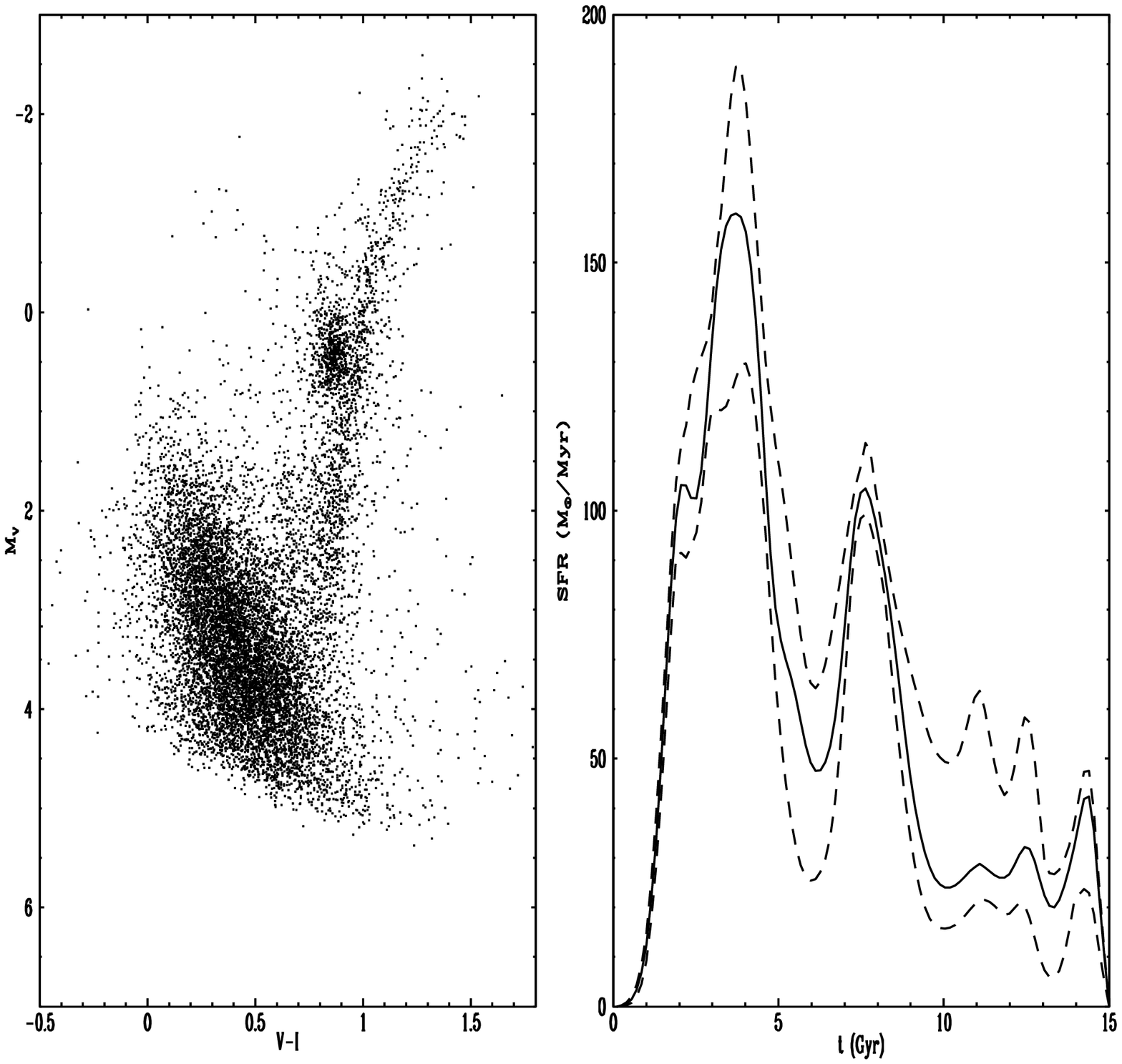,angle=0,width=18.1cm,height=9.0cm}
\@ \textbf{Figure 4.}\hspace{5pt}{
\begin{flushleft}\textbf{Left:}  Observational HR diagram for Leo I. 
\textbf{Right:}  Inferred $SFR(t)$ for the central values of the observational parameters, solid 
line. The dashed curves represent the error envelope as defined by the quoted uncertainties in 
the foreground extinction and distance modulus.
\end{flushleft}}
\end{figure*} 

The observations we obtained for LeoI are shown in the CMD in the left panel of Figure(4). This 
contains 11334 stars, which was reduced to 8691 after exclusion of unsuitable stars (e.g. the red 
clump), as described for the previous two galaxies. The excluded region comprised stars with 
$M_{V} <1$ and $B-V > 0.7$ and $B-V <0.9$, though our results are highly insensitive to the details
of these cut, as the RGB close to the red clump is an age degenerate region. The CMD of this 
galaxy reveals a young MS, but also a RGB extending down into the turn off region of a much older 
MS. The distribution of stars along the MS region is not uniform, and is actually encoding an 
interesting $SFR(t)$. 

Taking for this galaxy the central values of those given by Mateo (1998), $[Fe/H]=-1.5 \pm0.4, 
M_{V}=-11.9, E(B-V)=0.01\pm0.01$ and $(m-M)_{0}=21.99 \pm0.2$ (Reid \& Mould 1991, Lee et al. 
1993b and Demers et al. 1994), we invert the CMD of LeoI to obtain 
the $SFR(t)$ shown by the solid curve in the right panel of Figure (4). The dotted curves contain 
all other possible answers compatible with our method and the observational parameters taken with 
their errors. 

The $SFR(t)$ of LeoI can be divided into three distinct phases, which join continuously with no 
evidence of a discrete bursting behaviour. The first of these phases lasted from $15-10 Gyr$ ago, 
and proceeded at a rate of around $30 M_{\odot}/Myr$. The following two phases were extended peaks 
of star formation activity centered on ages of $8$ and $4 Gyr$, and having durations and maximum 
amplitudes of around $3$ and $4 Gyr$, at 100 and 150 $M_{\odot}/Myr$ respectively, as shown in 
Figure (4). Any total cessation of the star forming activity can be excluded for ages between
1 and 10 Gyr. As time resolution is lost beyond this age, the population beyond 10 Gyr could in 
principle be a single burst, and appear extended because of the observational errors.

The ground based study of Lee et al. (1993) reaching only the youngest turn off points, and 
subsequent analysis of this data set by Caputo et al. (1995) and Caputo et al. (1996) using 
isochrone matching techniques and luminosity function methods developed for single age globular 
clusters, revealed the presence of stars of ages 1-3 Gyr. Using more recent HST data and comparing 
to modeled CMDs Gallart et al. (1998) describe the star formation history of LeoI as coming mostly 
from an episode lasting from 6-2 Gyr ago, with the addition of an older component of duration 
2-3 Gyr, in good agreement with our inferred $SFR(t)$ for this galaxy. The population box of this 
galaxy given by Mateo (1998) is consistent with our results.

In this case, the $W$ test gives results showing that our inferred $SFR(t)$ is incompatible with 
the data at a two $\sigma$ level. As our inversion method has been extensively tested using 
synthetic CMD's, this result shows the data to be in conflict with our input assumptions. This 
is perhaps not surprising as this system has a much larger internal metallicity spread ($0.3 \pm 
0.1$) than the two reviewed previously, which together with the errors in the metallicity
determination allow for quite a large ($\sim 1$) internal spread. We have thus solved for the 
best fitting single metallicity solution, and discovered that internal metallicity dispersions 
are important. This metallicity spread introduces a time uncertainty going from 1-3 Gyr, for ages 
going from 1-13 Gyr. Solving simultaneously for the enrichment and star formation histories is a 
problem we shall treat later, as an extension of the variational calculus approach. A further 
possible source of the disagreement found between the synthetic CMD reproductions of our recovered 
$SFR(t)$ and the data is the difficulty of modeling precisely the error structure present in real 
data, as pointed out by Aparicio \& Gallart (1995). Although the inversion method itself is highly 
robust to the details of the assumed error structure, the very careful comparison of the $W$ test
would pick up any discrepancy between the assumed error structure used in generating the synthetic
CMD's, and that actually present in the data. Finally, the presently available isochrones do not 
take into account the relative overabundance of $\alpha$ elements at low metallicities, which at 
some level introduces a slight mismatch between the observed stars and the assumed modeling. 
Unfortunately, we can not distinguish between these possibilities easily.

\begin{figure*}
\epsfig{file=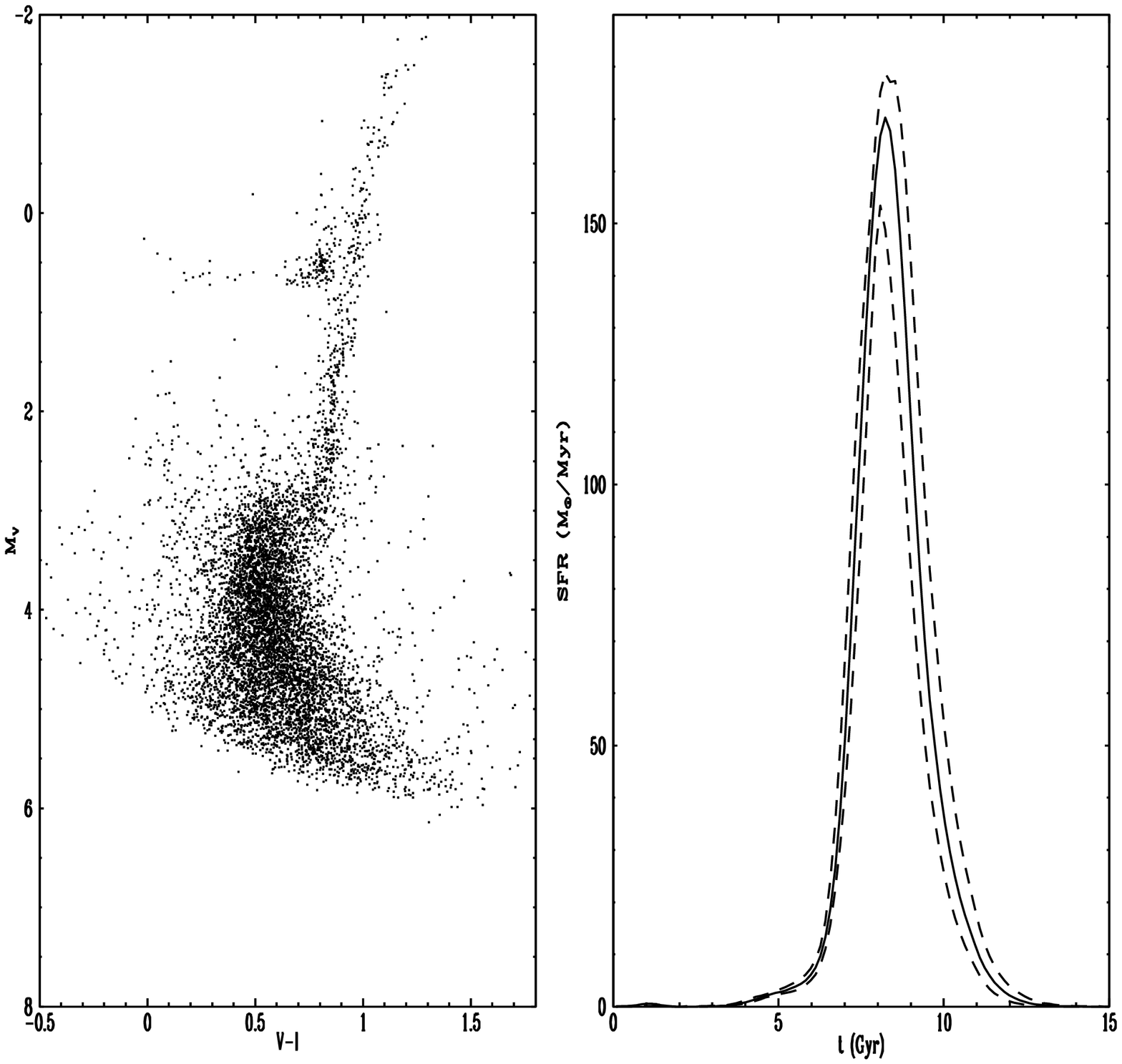,angle=0,width=18.1cm,height=9.0cm}
\@ \textbf{Figure 5.}\hspace{5pt}{
\begin{flushleft}\textbf{Left:}  Observational HR diagram for Leo II. 
\textbf{Right:}  Inferred $SFR(t)$ for the central values of the observational parameters, solid 
line. The dashed curves represent the error envelope as defined by the quoted uncertainties in 
the foreground extinction and distance modulus.
\end{flushleft}}
\end{figure*} 

\subsection{LeoII}

For this galaxy we obtained 7625 stars, of which we used 4492 after removing from the analysis 
the red clump (stars with both $M_{V}<1$ and $B-V <0.9$), the lower regions and some blue 
stragglers blue wards of $V-I=0.2$, which do not 
correspond to any of the evolutionary phases included in our isochrones. The full observational 
CMD is shown in the left panel of Figure (5). For the central values of the observational 
parameters of this galaxy as summarized by Mateo (1998) of $[Fe/H]=-1.9 \pm0.1, M_{V}=-8.9, 
E(B-V)=0.02\pm0.01$ and $(m-M)_{0}=21.63 \pm0.09$ (Mighel \& Rich 1996, Demers \& Irwin 1993
and Lee 1995) we obtain a divergent solution, indicating that 
the isochrones used do not correspond to the stars being analyzed. Changing the metallicity to 
$[Fe/H]=-1.75$, just marginally outside the errors reported by Mateo (1998), gives a stable 
convergence of the method and a significant result. The recent study by  Mighell and Rich (1996)
determined a metallicity of  $[Fe/H]=-1.6 \pm 0.25$ for this galaxy, consistent with what was 
used here. Our inferred $SFR(t)$ is shown by the solid curve in the right panel of Figure (5). 
Again, the dotted curves represent an envelope to all alternative reconstructions obtained by 
varying the observational parameters within their errors. As with LeoI, this galaxy shows a large 
internal metallicity spread, which is probably what the very sensitive $W$ test detects, also
giving an incompatible result between the model-model and data-model comparisons at more than a 
two $\sigma$ level. The discussion of this point given for the previous galaxy applies also to 
LeoII, metallicity spreads will have to be considered for a more accurate rendering of the star 
formation history of this galaxy.

In this case, we see a gradually rising $SFR$ from 12 Gyr to a peak of 160 $M_{\odot}/Myr$ at 8 
Gyr, followed by a somewhat more abrupt descent, with star formation activity ending by around 
6 Gyr ago, as shown by Figure (5). This result would be affected by the internal metallicity 
spread of $0.3$ dex of LeoII (Mateo 1998), producing a broadening of around 1-3 Gyr, see Table 
(1). Comparing with the study of Mighell and Rich (1996), who analyze an HST CMD of LeoII by 
fitting a ``fiducial sequence'' to the CMD and then comparing it to theoretical isochrones to 
solve for the age of the galaxy treated as a single parameter, we find no inconsistencies. They 
obtain an age of $9\pm 1$ Gyr for LeoII, with an age spread of around 4 Gyr, which is compatible 
with our results. They also report some degree of star formation at ages $>10 Gyr$, of which we 
see no evidence. This discrepancy is probably the result of the different methods used in the 
analysis. Given the high age resolution of our method, it is not only the median age and a 
representative value for the spread that we obtain, also the shape of the burst is recovered, 
ruling out for example a rectangular burst for this galaxy. Not just the age and duration of 
star formation episodes in these galaxies, but also the time structure of them can now be 
reliably inferred, and used in aiding theoretical interpretations of the origin of these systems.

\subsection{Draco}

\begin{figure*}
\epsfig{file=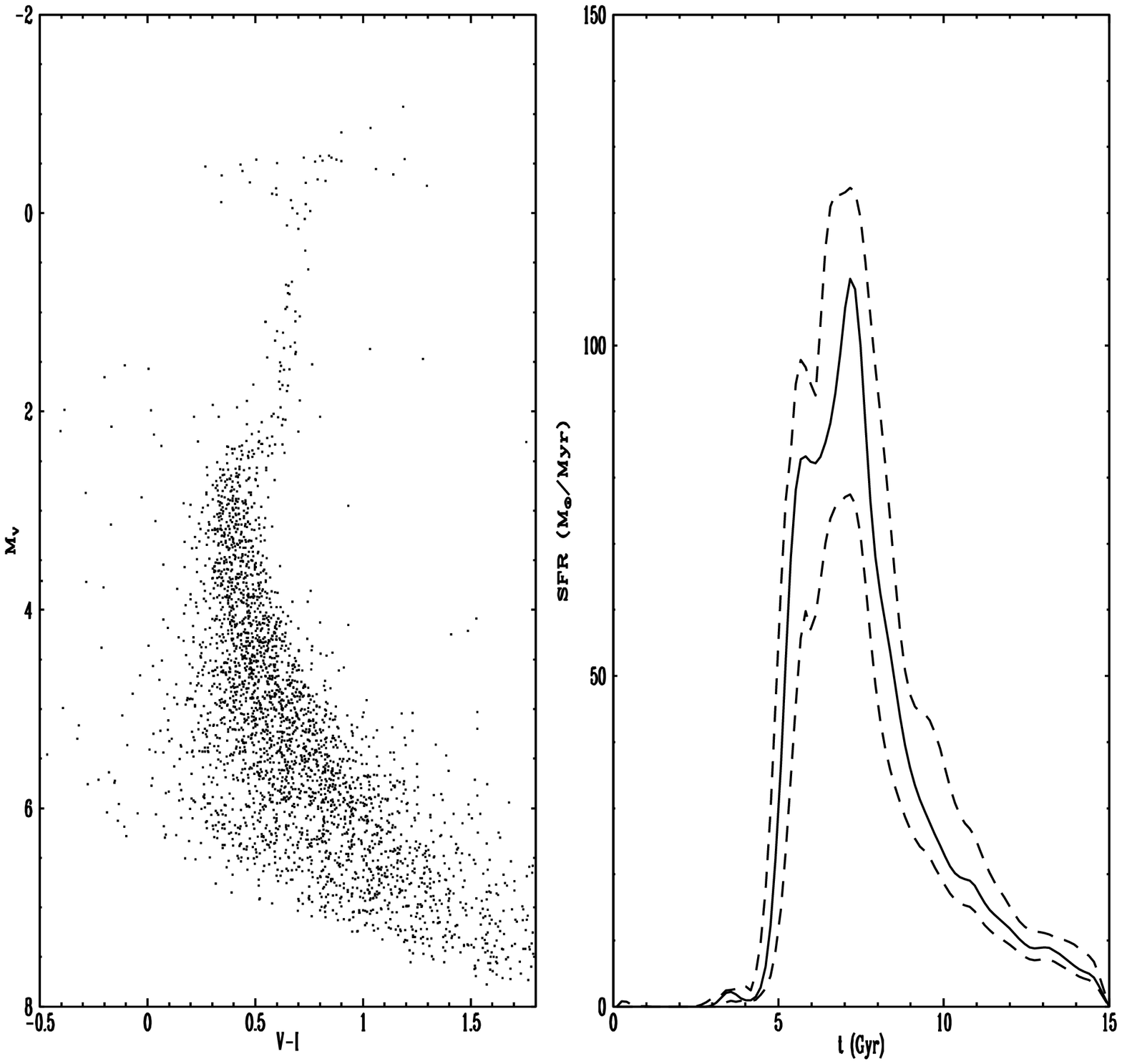,angle=0,width=18.1cm,height=9.0cm}
\@ \textbf{Figure 6.}\hspace{5pt}{
\begin{flushleft}\textbf{Left:}  Observational HR diagram for Draco. 
\textbf{Right:}  Inferred $SFR(t)$ for the central values of the observational parameters, 
solid line. The dashed curves represent the error envelope as defined by the quoted uncertainties 
in the foreground extinction and distance modulus. 
\end{flushleft}}
\end{figure*} 

For this last galaxy our observational CMD contains 3091 stars, most of which are located in the 
lower, age-degenerate region of the diagram, and were thus excluded from the analysis, leaving 
1210 stars after removing also the horizontal branch and blue stragglers (stars with both $M_{V} 
<0$ and $B-V<0.7$). Our full observational 
CMD for Draco is shown in the left panel of Figure (6). Mateo (1998) summarizes the metallicity 
and observational properties as $[Fe/H]=-2.0 \pm0.15, M_{V}=-8.8, E(B-V)=0.03\pm0.01$ and 
$(m-M)_{0}=19.58 \pm0.15$ (Carney \& Seitzer 1986, Lehnert et al. 1992, Nemec 1985 and Grillmair 
et al. 1998). Using the central values we produced our observational CMD and inverted 
it to obtain the $SFR(t)$ plotted as the solid curve in the right panel of Figure (6). The dotted 
curves contain all variations obtained by shifting our observations within the error ranges of the 
values given by Mateo (1998) for $E(V-I)$ and $(m-M)_V$.

Our result for Draco is very similar to what we obtained for LeoII, only shifted by 2 Gyr towards 
younger ages, giving a median age of around 7 Gyr. The time structure of the $SFR(t)$ also differs 
slightly from that of LeoII in that the peak is broader in Draco, having a plateau lasting around 
2 Gyr, rather than a narrow maximum. The presence of a low level old component extending beyond 
10 Gyr is also inferred, although the precise time structure in this region is not well restricted 
by the method. The values given by the normalization through the total luminosity are in the range 
of our other galaxies, with the maximum rate being of around 110 $M_{\odot}/Myr$. Using Saha's $W$ 
test in this case gives $208 \pm9$ for the model-model comparison, and $201 \pm 8$ for the 
data-model comparison, which gives us confidence in our results, as it shows our $SFR(t)$ is 
compatible with the data at better than a 1 $\sigma$ level.

Comparing this result with the recent study of Grillmair et al. (1998) we find the two results to 
be only marginally consistent, as they report an age of 10-12 Gyr (for the IMF we assumed) for the 
bulk of the stellar population of Draco $\pm 2.5 Gyr$, which they identify as essentially a single 
age event. We note several differences between their approach and ours, any of which on its own 
could bring the two results into closer agreement. They use HST data to construct their 
observational CMD, which is very similar to the one we obtain, no differences are evident at this 
level. Their analysis however differs markedly. They fit a fiducial sequence to the CMD diagram 
assumed to be representative of the bulk of the population, adjusting the MS and lower RGB regions 
through an inclusion envelope criterion, and the bright RGB by eye. This fiducial sequence is then 
compared to theoretical isochrones (VandenBerg \& Bell 1985) through a maximum likelihood analysis 
designed to find the age of the system, treated as a one parameter problem. We note that comparing 
any fiducial sequence to theoretical isochrones will only be a meaningful statistical procedure in 
cases where the underlying $SFR(t)$ is indeed a single epoch burst, e.g. in the case of a globular 
cluster. Further, defining any such sequence so that it is a valid statistical representation of 
the underlying $SFR(t)$ is a problem that has not been treated yet. Grillmair et al. (1998) also 
noted their isochrones showed systematic inconsistencies when compared to the stars they were 
dating, which also casts some doubt on their results, as they remarked. They also had the 
difficulty of requiring multiple conversions between their observational bands and those available 
from theoretical stellar models. 

Our results for this galaxy are weakened by the assumption of a single metallicity 
for the entire population. Although this can not be rigorously correct, it essentially holds for 
the previous four galaxies, which show small internal metallicity spreads. Draco however, has an 
internal spread of 0.5 dex (Mateo 1998), which could alter our results for this galaxy at the level
of 1-3 Gyr. Another possible explanation to the difference between our results and those of 
Grillmair et al. (1998) is that we used the Padova isochrones (Fagotto et al. 1994, Girardi et al. 
1996) rather than those of VandenBerg \& Bell (1985). Finally, we note that Carney \& Seitzer 
(1986) sampled a much larger region of Draco using ground based CCD data, and detected multiple 
turnoffs in this galaxy, corresponding to ages between 8 and 15 Gyr.

\section{Summary}

We have used a homogeneous set of observational colour magnitude data to study the star formation
histories of a sample of 5 dSph
galaxies, through a non-parametric variational calculus maximum likelihood method. We then performed a detailed
statistical analysis to check the accuracy of our results for each galaxy, obtaining good results
for three of our galaxies (Carina, Ursa Minor and Draco), and evidence of a systematic difference between 
our data and results for LeoI and LeoII, probably due to internal metallicity spreads. We can now
compare the results we obtained for the different galaxies, with the added consideration of a possible extra 1-3 Gyr
error margin in the results for LeoI and LeoII.

Ursa Minor appears to be the only essentially ``Population II'' system, being characterized by
a uniformly old star formation history. LeoII and Draco are systems which show 
similar star formation histories, being basically characterized by a single major episode. This lasted
in both cases around 4 Gyr centered at 8 Gyr, although
Draco shows a low level extension into much older ages. LeoI
shows the most complex $SFR(t)$, having a small old component of age $>10$ Gyr, and two later episodes
centered at 8 and 3.5 Gyr, although the star formation activity did not stop altogether between them. 
It is interesting that the second episode in LeoI coincides in age with the period of star 
formation in Draco and LeoII. Finally, star formation in Carina highly resembles that in LeoI in the relative
amplitude, duration and locations of the two main components. 

Since we used the total luminosities of these galaxies to normalize the inferred $SFR(t)$ in physical
units, we can derive other quantities of interest, for example the supernova rates as a function
of time. The SN type II rates are obtained by scaling the total $SFR(t)$ by a factor given by the fraction
of stars more massive than $8 M_\odot$, which for our assumed IMF translates 
$100 M_{\odot}/Myr \approx 1 SNII/2 Myr$. The only galaxy showing rates greater than $150 M_{\odot}/My$ 
is Ursa Minor, and it is also the only one with a $SFR(t)$ consistent with a single epoch burst.
The other four systems, showing extended $SFR(t)'s$, have rates of always less than 
$150 M_{\odot}/Myr$.
This last fact might indicate the presence of a threshold in these systems, above which energy input from
massive stars into the gas component is sufficient to totally disrupt the galaxies interstellar medium
 and end star formation.
No characteristic timescales of $\sim 1$ Gyr are evident from the recovered
$SFR(t)'s$ of these galaxies, with the possible exception of Ursa Minor, suggesting that 
SN type I are not determinant in driving the star formation processes in these systems.

Figure (7) presents the luminosity weighted sum of the
$SFR(t)$ for all the galaxies we analyzed. This shows the star formation activity in the set of these
low metallicity dSph's to have ended by around 2 Gyr ago, and having been relatively 
steady during the period 3-9 Gyr. 
For older ages we see the average $SFR(t)$ to be essentially dominated by the old age Ursa Minor galaxy.
Our results as summarized by Figure (7) support the calculations presented by Unavane et al. (1996) in that
the average metal-poor dSph star is of intermediate age, and not as old as a ``Population II'' 
halo star. 
A more complete sample would include the Sagittarius dwarf, with a mean age of $\approx 10 Gyr$, as well as the
much larger Magellanic clouds, having ages of $<3 Gyr$. It seems reasonable to suppose the total
star formation history for the satellites of the Milky Way to show no preferred epoch of star formation,
as suggested by Tolstoy (1998).

From these comparisons it is clear that the dSph galaxies of the Milky Way do not form a simple
system, and straight forward correlations between $SFR(t)$ and other present day parameters such as
instantaneous galactocentric distance or metallicity are not evident, and perhaps not even meaningful.
A physical understanding of these systems will probably have to consider the complex interactions of
these systems with the halo of the Galaxy (tidal forces, evolving gaseous component, 
orbital structure etc). It could well be the case that the present day sample of survivors actually experienced
very distinct origins and evolutionary histories (as direct studies of their $SFR(t)$ appear to indicate) 
with little in common other than having been shaped under the dominating influence of the Milky Way.
A ram pressure stripped dwarf irregular and a more recently tidally torn fragment of the Magellanic clouds
could both end up as dSph systems today.

\begin{figure}
\epsfig{file=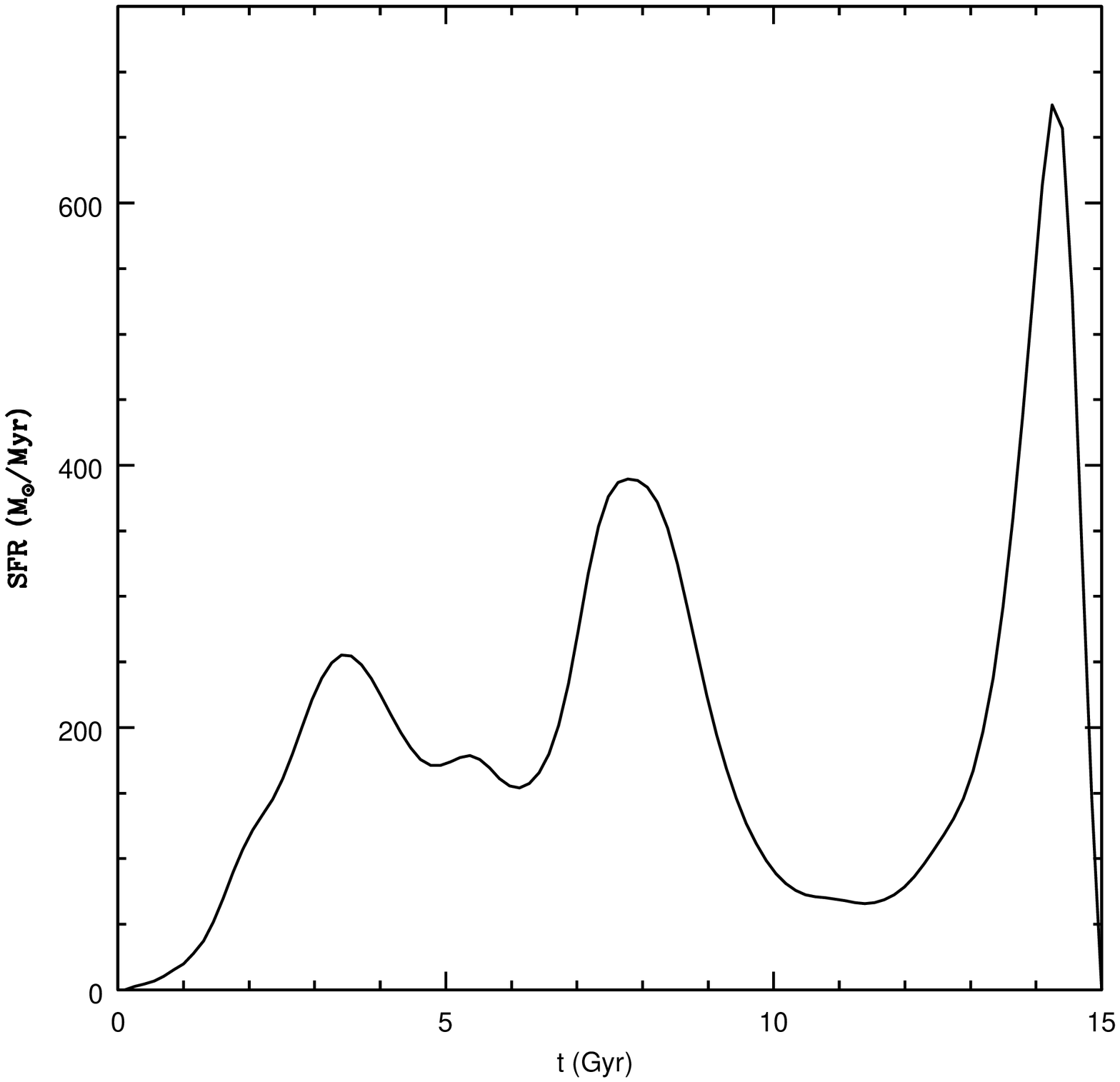,angle=0,width=8.5cm}
\@ \textbf{Figure 7.}\hspace{5pt}{Sum of the $SFR(t)'s$ of the five galaxies we studied}
\end{figure}

Much more information is needed on these systems before their full potential as tracers of the build up
and evolution of the Milky Way can be realized, once their individual evolutionary histories are better
understood. Studies aimed at recovering the full orbital structure of these galaxies, through proper
motion measurements and potential theory reconstructions will yield crucial independent information
on the evolution and formation of these systems. A more complete sampling than the one we have conducted
here is needed to visualize local dSph's fully. Any advance in the modeling of advanced stellar
phases will also improve the use of CMD as tools in galactic evolution studies, as it would in principle
eliminate the need to remove parts of the CMD from consideration. 

Our present sample includes the 5 dSph galaxies having the lowest internal metallicity spreads, and 
therefore the ones for which our present method applies best. Obtaining a larger sample and 
analyzing it through a fully consistent and non parametric statistical method, will require the 
simultaneous recovering of the enrichment history and the $SFR(t)$. Development of such a method
will be the subject of a future work. We emphasize the need of a homogeneous
sample at all levels of the analysis, together with a fully consistent statistical inversion method 
which does not assume any {\it a priori} structure for the $SFR(t)$ one is solving for, in 
comparative studies of star formation histories.

\section*{Acknowledgments}

The work of X. Hernandez was partly supported by a DGAPA-UNAM grant.

\appendix
\section{The images}

The images were recovered from the HST data archive. The image numbers, filters
and observation dates are given in table A1. Table A2 gives the A-to-D gains, exposure 
times, and numbers of images for the dSph fields.  

\begin{table}
\centering
\caption{The dwarf spheroidal galaxy data used. The galaxy name, the date of
observation, the HST archive image numbers and the filters used are given}
\begin{tabular}{ccccc}
\\
Galaxy & Date of     & Image     & \multicolumn{2}{c}{Filters Used} \\
Name   & Observation & Filenames & `V'-filter & `I'-filter          \\
\\
    Leo II & 15 May 1994 & u28v0101t-kt & F555W & F814W \\
     Leo I &  5 Mar 1994 & u27ko101t-8t & F555W & F814W \\
     Draco &  9 Jun 1995 & u2oco101t-6t & F606W & F814W \\
    Carina &  3 Jan 1995 & u21b0101t-6t & F555W & F814W \\
Ursa Minor &  4 Jul 1995 & u2pb0101t-6t & F555W & F814W \\
\\
\end{tabular}
\label{chap6tab1}
\end{table}

\begin{table}
\centering
\caption{The A-to-D gains, exposure times, and numbers of images for the dSph
fields.}
\begin{tabular}{cccc}
\\
Galaxy & A-to-D & \multicolumn{2}{c}{Total exposure (secs)}\\
Name & gain & `V'-filter & `I'-filter \\
\\
    Leo II & 7.0 & 4800 & 4800  \\
     Leo I & 7.0 & 5700 & 4800  \\
     Draco & 7.0 & 2000 & 2400  \\
    Carina & 7.0 & 2200 & 2200  \\
Ursa Minor & 7.0 & 2100 & 2300  \\
\\
\end{tabular}
\label{chap6tab3}
\end{table}

\subsection{Retrieving the data and image preparation}

For each of the five dSph's, there were a number, $n$, of images
taken, in both a long band (F814W, corresponding closely to Johnson's
I), and a short band (F555W or F606W, corresponding to Johnson's V).

For each dSph, there are a set of long exposure `V' and `I' data
files. Each data file contains the output from 4 detectors -- the
planetary camera (PC), and the three Wide Field cameras (WF2, WF3 and
WF4). Each of these detector images is 800$\times$800 pixels in size,
of which typically 730$\times$730 are usable.

Data treatment was carried out in the IRAF environment. Image
combination was carried out using the STSDAS package, and photometry
using the DAOPHOT package.

In order to remove the severe cosmic ray effects in each image, the
$n$ images were combined by taking the mean value for each pixel
position, after the rejection of values either too high or too low
with respect to local variations.  The task used here was `crrej' in
the STSDAS package. Each image was also cropped to a 730$\times$730
image by rejecting the first 60 rows/columns and the last 10.

In the following analyses, only the data from the three WF cameras
were used.  The pixel scale in these WF images is 0.10$"$ per pixel.

\subsection{Source extraction}

Sources were extracted to 2$\sigma$ above the mean background. Point
spread function (PSF) fitting photometry was carried out by selecting
isolated stars to define a PSF. Sources with a fit at $\chi^2 > 1.5$
were rejected, and the magnitudes used were aperture magnitudes with a
radius of 2 pixels. (0.20$"$).

\subsection{Galactic contamination}

At faint magnitudes in optical wavebands, external galaxies can
constitute a major contaminant in number counts. Williams et
al. (1996), based on HDF (Hubble Deep Field) galaxy counts find
$\sim$2$\times$10$^5$ galaxies per square degree brighter than
V=26. For the area of the three WF detectors, this corresponds to, on
average, 250 contaminating galaxies in the field of view. Compared
with the many thousands of stars in the images, this contamination is
small.

Nevertheless, the high resolution of the HST allows the separation of
galaxies from stars more reliably than for ground based work, where
image resolution is necessarily lower and the distinguishing stars from
galaxies is less straightforward.

The cut in $\chi^2$ which we use eliminates the majority of galaxies
(and remaining cosmic ray events) because these will in general be
fitted poorly by the PSF.

\subsection{Magnitude corrections}

\subsubsection{Aperture Corrections}

An aperture correction to render these magnitudes equivalent to those
which would be obtained by use of a 0.5$"$ radius aperture (see
Holtzman et al. 1995) was found by selecting bright, unsaturated
stars, in each of the `V' and `I' bands, and for each detector and
each dSph separately. The mean differences between the magnitude using
a 5.02 pixel diameter aperture (0.5$"$ radius) and a 2 pixel diameter
aperture were used to correct all magnitudes.

\subsubsection{A to D gain correction}

All the observations for the dSph's considered here were taken through
bay 4 (see Holtzman et al. 1995), which means that the
Analogue-to-Digital gain is only 7.0, rather than the standard value
of 14.0. Due to some unshared electronics, this necessitates a
different correction for each of the WF fields, as indicated below:

\bigskip

\begin{tabular}{cc}
Wide Field & $\Delta$m \\
\\
2 & 0.754 \\
3 & 0.756 \\
4 & 0.728 \\
\\
\end{tabular}

\subsubsection{Geometric correction}

The WF cameras have geometric distortions which arise mainly from
elements in the optical path. As a consequence, the effective pixel
areas, in square angular measure, vary systematically across each WF
detector. We make a parameterization of the data from the figures of
Holtzman et al. (1995), and apply that as a correction. The
correction is well represented by a quadratic function of distance
from the centre of the detector, and never exceeds 0.04 magnitudes at
the edge of the detectors.

We use $\Delta m = -1.897 \times 10^{-5}$ $r - 1.208 \times 10^{-7}$ $r^2$ 
where $r$ is the
distance in pixels from the position (400,400) on the detector.

\subsubsection{Charge Transfer efficiency correction}

The readout of the CCD detectors requires the transfer of charge
through successive rows of the detector. As a consequence, the signal
from the last rows to be read are diminished because of the loss of
charge during transfers. The correction is a maximum of 0.04
magnitudes at the final row.

We use $\Delta m = -0.04 (y/800)$ where $y$ is the row number.

\subsubsection{Corrections for Leo I}

WFPC2 data taken before 23rd April 1994 was taken at a detector
temperature of $-$76$^{\circ}$C rather than $-$88$^{\circ}$C. The
change reduced CTE effects, and IR zeropoint problems. The Leo I
observations were made before this change, and we use a linear ramp of
size 0.12 magnitudes to correct for the CTE effects as recommended by
Holtzman et al. (1995), and additionally, an additive offset of 0.05
magnitudes in I to correct for the higher zero point.

\subsection{Conversion to standard V and I}

Reddening must be taken into account before corrections are made to V
and I. We use the published values of reddening, which are fairly
small for all these galaxies ($E(B-V) < 0.08$), to make corrections to
the magnitudes before applying the transformations given below.

We use the extinction values tabulated by Holtzman et al. (1995) for
the filter F555W and F814W, and an estimate based on these for the
F606W filter:

$ A_{F555W} = 3.026 E(B-V) $

$ A_{F814W} = 1.825 E(B-V) $

$ A_{F606W} = 2.75 E(B-V) $ (estimate)

We subsequently apply synthetic transformations given by Holtzman et
al. (1995), of the form
 
\medskip

Output Band = $m_{raw}$ + $a_0$ + $a_1$(V$-$I) + $a_2$(V$-$I)$^2$

\medskip

where the coefficients are given in table \ref{holtztrans}.

\begin{table}
\centering
\caption{The synthetic transformations from F555W and 
F606W to V, and from F814W to I, taken from Holtzman et al. (1995)}
\bigskip
\begin{tabular}{lccccc}
\hline
Filter & Constraint & Output Band  & $a_0$ & $a_1$ & $a_2$ \\
\hline
F555W &    --         & V & 21.729 & $-$0.051 & 0.009 \\
F606W & (V$-$I)$<$1.0 & V & 22.093 & 0.254 & 0.012 \\
F606W & (V$-$I)$>$1.0 & V & 22.883 & $-$0.247 & 0.065 \\
F814W & (V$-$I)$<$1.0 & I & 20.838 & $-$0.012 & $-$0.006 \\
F814W & (V$-$I)$>$1.0 & I & 20.920 & 0.028 & $-$0.124 \\
\hline
\end{tabular}
\label{holtztrans}
\end{table}

$m_{raw}$ is the output aperture magnitude using an aperture of 2 pixels
radius, corrected as indicated above to an aperture of radius 0.5$"$.

The above formulae were iterated until no further significant change
in V or I occurred.

Finally, the reddenings which had been removed are restored using:

$ A_V = 3.10 E(B-V)$

and

$ A_I = 1.83 E(B-V)$

The reddenings used are refined iteratively after the fitting of
isochrones to the MS region. Note, however, that
because the reddenings are small, and the reddenings are restored to
the data afterwards, even if the initial guess for the reddening is
wrong by a substantial amount, very little effect is seen in the final
photometry. (e.g. a change of the reddening by 0.1$^m$ changes
photometry by less than 0.005$^m$.)

\subsection{Systematics}

As Holtzman et al. (1995) point out, there remain some aspects of the
photometric calibration of WFPC2 which are uncertain. For example,
Holtzman et al.(1995) note discrepancies of $\sim$ 0.05 magnitudes between
long and short exposures, which are not understood. Several more minor
systematic and random effects at the level of a few percent
(corresponding to a few hundredths of a magnitude) are not well
understood. Furthermore, the conversion that has been made here to
standard V and I colours, introduces more minor uncertainties. 
It must be noted that the systematic effects in zeropoints may be
as large as 0.1 magnitudes. Any such uncertainties appear as an offset
primarily in adopted distance modulus, and would affect most CMDs in
the same way. Differentially, any effects should be minimal.

\end{document}